\documentstyle[12pt,epsfig]{article}
\renewcommand{\a}{\alpha}
\renewcommand{\b}{\beta}

\newcommand{\la}{\lambda}
\newcommand{\G}{\Gamma}
\renewcommand{\S}{\Sigma}
\newcommand{\La}{\Lambda}

\newcommand{\bea}{\begin{eqnarray}}
\newcommand{\eea}{\end{eqnarray}}
\newcommand{\beq}{\begin{equation}}
\newcommand{\eeq}{\end{equation}}
\newcommand{\nn}{\nonumber}
\newcommand{\fr}{\frac}
\newcommand{\hl}{\hline}

\begin{document}

\begin{titlepage}
\begin{flushright} {FTUV/99-33 \\ IFIC/99-34}
\end{flushright}
\vskip 2cm
\centerline{\LARGE \bf
Magnetic Moments of  Heavy Baryons}
\vskip 1cm
\centerline{M.C. Ba\~nuls, I. Scimemi,}
\centerline{J. Bernab\'eu, V. Gim\'enez, A. Pich.}
\vskip 0.5cm
\centerline{Dep. de F\'{\i}sica Te\'orica, IFIC, Univ. de Valencia-CSIC,}
\centerline{ E-46100 Burjassot (Valencia), Spain}

\begin{abstract}

First non-trivial chiral corrections  to the  magnetic moments 
 of  triplet ($T$)  and  sextet ($S^{(*)}$) heavy baryons are calculated
using    Heavy Hadron Chiral Perturbation Theory.
Since  magnetic moments of
 the $T$-hadrons 
  vanish in the    limit of infinite heavy quark mass ($m_Q\rightarrow\infty$),
these corrections  occur
  at order ${\cal O}(1/(m_Q \Lambda_\chi^2))$ for $T$-baryons
while  for  $S^{(*)}$-baryons they are
 of order  ${\cal O}(1/\Lambda_\chi^2)$. 
The renormalization  of the chiral loops 
 is discussed  and 
 relations among  the magnetic moments of different hadrons  are provided. 
Previous results for $T$-baryons are revised.

\end{abstract}

\end{titlepage}
\newpage
\section{Introduction}

Chiral Perturbation Theory (ChPT) and Heavy Quark Effective Theory (HQET)
 can be combined together to construct an effective lagrangian which
 describes soft interactions of hadrons containing a single heavy quark~\cite{wise}-\cite{cho}.
Electromagnetic interactions can be included in the formalism 
by gauging a $U(1)_{EM}$ subgroup of the global $SU(3)_L \otimes SU(3)_R$ 
symmetry group.

In the limit $m_u$, $m_d$, $m_s \rightarrow 0$, the QCD Lagrangian for 
light quarks has a $SU(3)_L \otimes SU(3)_R \otimes U(1)_V$ symmetry, 
which is spontaneously broken to $SU(3)_V \otimes U(1)_V$.
The lightest particles of the hadronic spectrum, the pseudoscalar 
octet ($\pi$, $K$, $\bar{K}$, $\eta$), can be identified with the 
corresponding Goldstone bosons.
Their low-energy interactions can be analysed making use of Chiral 
Perturbation Theory (ChPT) \cite{pich}, which is an expansion in 
terms of momenta and meson masses. 
Goldstone bosons are parametrised as
\beq\label{pi}
{\bf \pi}= \frac{1}{\sqrt{2}} \left ( \begin{array}{ccc}
\sqrt{\frac{1}{2}} \pi^0 + \sqrt{\frac{1}{6}} \eta & \pi^+ & K^+ \\
\pi^- & - \sqrt{\frac{1}{2}} \pi^0 +\sqrt{\frac{1}{6}} \eta & K^0 \\
K^- & \bar{K}^0 & -\sqrt{\frac{2}{3}} \eta \end{array}
\right ),
\eeq
and appear in the Lagrangian via the exponential representation 
$\Sigma=\exp(2 i {\bf\pi} /f_{\pi})\equiv \xi^2$,  being $f_\pi\sim 93$ MeV the 
pion decay constant.
Under chiral transformations,
\beq
\Sigma \rightarrow L \Sigma R^+ \quad \xi \rightarrow L \xi h^+ = h \xi R^+,
\eeq
where $L (R)$ are global elements of $SU(3)_{L(R)}$, and $h$ is a local 
$SU(3)_{L+R}$ transformation,
 which depends both on $\pi$ and on the chiral trasformation $(L,\ R)$.
To construct the 
effective theory, one must write the most general 
Lagrangian consistent with chiral symmetry involving $\Sigma$ and its derivatives.
Chiral symmetry is explicitly broken in QCD by the quark mass term.
This can be incorporated in the effective Lagrangian through 
the light mass matrix $\chi$, which gives rise to a quadratic 
pseudoscalar-mass term.
The Lagrangian is then organized as
an expansion in powers of $(p/\Lambda_{\chi})$ and $(m_{q}/\Lambda_{\chi})$,
where $p$ is the low pseudoscalar momenta, $m_{q}$ denotes the light quark masses
and $\Lambda_{\chi}\approx 1$ GeV is the chiral symmetry breaking scale
which suppresses higher--order  terms in the effective theory.

On the other side, in the opposite limit, $m_Q \rightarrow \infty$, 
which is useful for $c$ and $b$-quarks, different simplifications 
occur in the dynamics of heavy-light hadrons.
Quark interactions do not change the velocity of the heavy quark 
inside the hadron, because the momentum exchange is of order 
$\delta P \sim \Lambda_{QCD}<\!< m_Q$.
In the hadron rest frame, the heavy quark acts as a static colour 
source which interacts with the light degrees of freedom.
This interaction is independent of the mass and spin of the heavy 
quark, and thus the hadron dynamics shows $SU(2)$ spin symmetry 
and $SU(N_f)$ flavour symmetry (for $N_f$ heavy flavours).
HQET~\cite{pichLH}
 is an effective field theory for QCD which 
makes this symmetry manifest in the  $m_Q \rightarrow \infty$ 
limit, and describes the dynamics of hadrons containing a heavy 
quark, at momenta much lower than $m_Q$.
 The effective baryon fields are labeled by their velocities and 
their mass is removed from the baryon momentum, $P$.
Derivatives on the baryon fields produce powers
 of residual momentum $(k/m_Q)<< (P/m_Q)$.

In some kinematical regions, which are not far from the chiral 
and heavy quark limits, both approaches can be simultaneously used. 
Baryons containing a heavy quark, 
in the $m_{Q}\rightarrow \infty$ limit, can emit and absorb light 
pseudoscalar mesons without changing its velocity, $v$. 
In Heavy Hadron Chiral Perturbation Theory (HHCPT) one constructs 
an effective Lagrangian whose basic fields are heavy hadrons and light mesons.
In ref.~\cite{chogeo},
 the formalism is extended to include also electromagnetism.
We use this hybrid effective Lagrangian to calculate 
the magnetic moments (MM)
of some baryons containing a $c$ or a $b$ quark.

In section~\ref{sec:form} we review the needed
HHCPT formalism, introduced in ref.~\cite{chogeo}:
the effective fields representing $S$ and $T$-baryons, 
the lowest order chiral lagrangian and its ${\cal O}(1/m_Q)$,
${\cal O}(1/\La_\chi)$ corrections.
These terms generate divergent chiral loops which contribute to the MM.
Their renormalization  requires the introduction of 
higher order operators.
In the case of $S$ we find that all divergences and scale dependence 
to ${\cal O}(1/\Lambda_\chi^2)$ 
can be absorbed in a redefinition of  only one  ${\cal O}(1/\Lambda_\chi)$
coupling.
Our computations and results are presented in section~\ref{sec:S}.
 The magnetic  moments of the $T$-baryons are analyzed 
in ref.~\cite{sava}. 
However this analysis does not include all meson loops and the needed 
counterterms are not taken into account.
In section~\ref{sec:T} we provide a consistent calculation of the $T$ magnetic moments to order 
 ${\cal O}(1/(m_Q \Lambda_\chi^2))$.
Finally section~\ref{sec:fin} summarizes our conclusions.

\section{HHCPT formalism for magnetic moments}
\label{sec:form}

The light degrees of freedom in the ground  state  of a 
baryon  with one heavy quark can be either in
a $s_l=0$ or in a $s_l=1$ configuration.
The first one corresponds  to $J^P=\frac{1}{2}^+$ baryons, which 
are annihilated by $T_i(v)$ fields which transform
as a $\bar{\bf3}$ under the chiral $SU(3)_{L+R}$ and as a doublet under 
the HQET $SU(2)_{v}$.
In the second case, $s_l=1$,
the  spin of the heavy quark and the light
 degrees of freedom combine together to form
$J^P=3/2^+$ and $J=1/2^+$ baryons which are degenerate
in  mass  in the $m_Q\rightarrow \infty$ limit.
 The spin-$\frac{3}{2}$ ones are annihilated
by the Rarita-Schwinger field $S_{\mu}^{* ij}(v)$ while the spin-$\frac{1}{2}$
baryons are destroyed by the Dirac field $S^{ ij}(v)$. It is very useful to combine
both operators into the so-called superfield \cite{falk} 
\bea
S_{\mu}^{ij} (v)&=& \sqrt{\frac{1}{3}} (\gamma_{\mu} + v_{\mu}) \gamma^5
                     S^{ij} (v) +  S_{\mu}^{* ij}(v) \ ,\nonumber \\
\bar{S}_{ij}^{\mu} (v) &=& - \sqrt{\frac{1}{3}} \bar{S}_{ij}(v) \gamma^5 
                       (\gamma^{\mu} + v^{\mu})+ \bar{S}_{ij}^{* \mu}(v) \ ,
\eea
which transforms as a {\bf 6} under
$SU(3)_{L+R}$ and as a doublet under $SU(2)_{v}$ and is symmetric in the $i$, $j$ indices.

The particle assignement  for the  $J= 1/2$ charmed baryons of 
the $\bar{\bf 3}$ and {\bf 6} representations is
\bea
(T_1,T_2,T_3)&=&(\Xi^0_c,-\Xi^+_c,\Lambda^+_c) \ ,
\\
S^{i j} &= &\left ( \begin{array}{ccc} 
\Sigma^{++}_c& \sqrt{\frac{1}{2}} \Sigma^+_c & \sqrt{\frac{1}{2}} \Xi^{+'}_c \\
\sqrt{\frac{1}{2}} \Sigma^{+}_c &\Sigma^0_c & \sqrt{\frac{1}{2}} \Xi^{0'}_c \\
\sqrt{\frac{1}{2}} \Xi^{+'}_c & \sqrt{\frac{1}{2}} \Xi^{0'}_c &
 \Omega^0_c 
\end{array}
\right )\ ,
\label{eq:sba}
\eea
and the corresponding $b$-baryons are
\bea
(T_1,T_2,T_3)&=&(\Xi^-_b,-\Xi^0_b,\Lambda^0_b) \ ,
\\
S^{i j} &= &\left ( \begin{array}{ccc} 
\Sigma^{+}_b& \sqrt{\frac{1}{2}} \Sigma^0_b & \sqrt{\frac{1}{2}} \Xi^{0'}_b \\
\sqrt{\frac{1}{2}} \Sigma^{0}_b &\Sigma^-_b & \sqrt{\frac{1}{2}} \Xi^{-'}_b \\
\sqrt{\frac{1}{2}} \Xi^{0'}_b & \sqrt{\frac{1}{2}} \Xi^{-'}_b &
 \Omega^-_b 
\end{array}
\right ) \ .
\label{eq:sbab}
\eea
The  $J=3/2$  partners  of the   baryons of Eq.~\ref{eq:sba} and 
Eq.~\ref{eq:sbab} have the same $SU(3)_V$ assignement in $S_\mu^{*ij}$.

The lowest order chiral lagrangian describing the soft hadronic and electromagnetic 
interactions of these baryons in the infinite heavy quark mass limit is given by \cite{chogeo}
\bea
{\cal L}^{(0)} &=& -i \bar{S}_{ij}^{\mu} (v \cdot D) S_{\mu}^{ij} +
 \Delta_{ST} \bar{S}_{ij}^{\mu} S_{\mu}^{ij} + i \bar{T}^i (v \cdot D) T_i \nonumber \\ 
&& + i g_2 \varepsilon_{\mu \nu \sigma \lambda} \bar{S}_{ik}^{\mu} v^{\nu} (\xi^{\sigma})_j^i
 (S^{\lambda})^{jk} \nonumber \\
&& +g_3 \left [\epsilon_{ijk} \bar{T}^i (\xi^{\mu})_l^j S_{\mu}^{kl}+ \epsilon^{ijk} \bar{S}_{kl}^{\mu} 
(\xi_{\mu})_j^l T_i\right ]\ .
\label{eq:lagos}
\eea
In this formula, the heavy baryon covariant derivatives are
\bea
D^{\mu} S_{\nu}^{ij}&=&\partial^{\mu}S_{\nu}^{ij}+(\G^{\mu})_k^i S_{\nu}^{kj}+
(\G^{\mu})_k^j S_{\nu}^{ik}-i e {\cal A}^{\mu}
 [Q_Q S_{\nu}^{ij}+Q_k^i S_{\nu}^{kj}+Q_k^j S_{\nu}^{ik}] \nonumber \\
D^{\mu} T_i &=& \partial^{\mu} T_i - T_j (\G^{\mu})_i^j - i e {\cal A}^{\mu} [Q_Q T_i -T_j Q_i^j] 
\ ,
\eea
where ${\cal A}^{\mu}$ is the electromagnetic current,
$Q_Q$ is the heavy quark  charge, the light quark charge
 matrix $Q$ is
\beq
{\bf Q} = \left ( \begin{array}{ccc}
\frac{2}{3} & & \\
& -\frac{1}{3} &  \\
& & -\frac{1}{3} 
\end{array} \right ) \ ,
\eeq
and the Goldstone fields appear through   axial-vector, $\xi_{\mu}$, and vector,
$\G_{\mu}$, currents 
\bea
 \xi_\mu &=& i(\xi D_{\mu} \xi^\dagger- \xi^\dagger D_{\mu} \xi)/2\nonumber\\
 \G_\mu &=& (\xi D_{\mu} \xi^\dagger+ \xi^\dagger D_{\mu} \xi)/2 \ ,\nonumber
\eea
with 
\beq
 D^{\mu}\xi = \partial^{\mu} \xi-i e {\cal A }^{\mu} [Q , \xi]\ . \nn
\eeq  
Because of the different spin configuration of the light degrees of freedom
 there is an intrinsic mass difference, $\Delta_{ST}\equiv M_S-M_T$,
 among the $S^{(*)}$  and the $T$ baryons.

Notice that a direct 
coupling  of the pseudo-Goldstone bosons to the $\bar{\bf 3}$ baryons is forbidden at the 
lowest order in $1/\La_\chi$.

As can be seen, there are no MM terms in the
lowest order lagrangian in Eq.~\ref{eq:lagos}. Therefore, the contributions
to the MM come from:\\
1) the next order in the baryon chiral lagrangian \cite{chogeo}
\bea
{\cal L}^{(long)} &=& 
\frac{e}{\Lambda_{\chi}}
 \left\{ i c_S \ {\rm tr }\left[\bar{S}_{\mu} Q S_{\nu} +\bar{S}_{\mu} S_{\nu} Q
\right] F^{\mu \nu} \right. \nn \\  
 & & + \left.   c_{ST}\
 [\epsilon_{ijk} \bar{T}^i 
v_{\mu} Q_l^j S_{\nu}^{kl} +
 \epsilon_{ijk} \bar{S}_{\nu,kl} v_{\mu} Q_j^l T_i ] \tilde{F}^{\mu \nu} \right \} \ ,
\label{eq:long}
\eea
where the constants $c_i$  are 
 all unknown.
We will take $\La_\chi =4\pi f_\pi\simeq 1.2$ GeV which fixes the normalization of 
 the unknown couplings $c_i$; \\
2) terms of order $1/m_{Q}$ from the heavy quark expansion which
break both spin and flavour symmetries \cite{chogeo}.
\bea
{\cal L}^{(short)} &=& -\frac{1}{2 m_Q} \bar{S}_{ij}^{\lambda} (i D)^2 S_{\lambda}^{ij}
 - \frac{e Q_Q}{4 m_Q}\bar{S}_{ij}^{\lambda} \sigma_{\mu \nu} S_{\lambda}^{ij} F^{\mu \nu}  \nn \\
& & 
 +\frac{1}{2 m_Q} \bar{T}^i (i D)^2 T_i + \frac{e Q_Q}{4 m_Q}\bar{T}^i \sigma_{\mu \nu}
 T_i F^{\mu \nu}\ ;
\label{eq:hqet}
\eea \\
3) chiral loops of Goldstone bosons coupled to photons, as described
by the lowest order Lagrangian.

\section{Results for  $S$-baryons ($s_l=1$)}
\label{sec:S}

\begin{figure}[ht]
\begin{center}
\epsfig{file=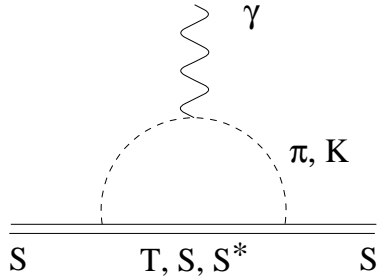,width=0.3\linewidth,angle=0}
\caption{Meson loops contributing to $S$-baryons MM.}
\label{fig:S}
\end{center}
\end{figure}

We  define the magnetic moment operator  for a spin 1/2 baryon $B$  and
 a spin 3/2 baryon $B^*_\nu$  respectively as
\bea
&-i e \mu(B)
F^{\a\b} \bar{B}\sigma_{\a \b}  B \ , &  \nn \\
&
-ie \mu(B^*)
 F^{\a\b} \bar{B^*_\mu}\sigma_{\a \b}  B^{*\mu} \ . & 
\label{eq:magop}
\eea
The  leading  contributions from the light- and heavy-quark magnetic interactions are of 
 order ${\cal O}(1/\La_\chi) $ and  ${\cal O}(1/m_Q) $ respectively.
We compute the next-to-leading chiral corrections of order 
${\cal O}(1/\La_\chi^2) $  which originate from the loop diagrams 
shown in fig.\ref{fig:S}.

The resulting MM can be decomposed  as:
\bea
\mu (B^{(*)})&=&\fr{1}{72} \left( 6 \fr{Q_{Q}}{m_Q} \mu_{HQE}(B^{(*)})+
 \fr{16 c_s}{\Lambda_\chi}\mu_{\chi}(B^{(*)}) \right. \nn \\
& & + \left.
3 g_2^{2} \fr{\Delta_{ST}}{(4 \pi f_\pi)^2}\mu_{g_2}(B^{(*)})-
3 g_3^{2} \fr{m_K}{4 \pi f_\pi^2}\mu_{g_3}(B^{(*)}) \right)\ .
\label{eq:magS}
\eea

where $  \mu_i (B)$ and $\mu_i(B^{*})$ are related by 
\beq
\begin{array}{cc}
\displaystyle{\fr{1}{3}}\ \mu_{HQE} (B^{*})=\ \mu_{HQE} (B) =1 & \nn \\
\mu_i (B^{*})=-\displaystyle{\fr{3}{2}}\ \mu_i (B) & i=\chi,\ g_2 ,\ g_3\ .
\end{array}
\eeq

\begin{table}
\begin{center}
\begin{tabular}{|c|c|} \hl
$f_\pi$ & 93 MeV \\
$m_\pi$ & 140 MeV \\
$m_K$ & 496.7 MeV \\
$\Delta_{ST}$ & 168 MeV \\
$m_c$ & 1.3 GeV \\
$m_b$ & 4.8 GeV \\
\hl
\end{tabular}
\caption{Constants used in  numerical estimates.}
\label{tab:cost}
\end{center}
\end{table} 

\begin{table}
\begin{center} 
\begin{tabular}{|c|c|c|c|c|}  \hl
 c quark  & b quark & 
 $ \mu_\chi  $   &
$\mu_{g_3}   $ &
$\mu_{g_2}   $ 
\\
\hl
$\Sigma^{++}_c$&$\Sigma^{+}_b$& 2&$1+m_\pi/m_K$   &$I_\pi$ + $I_K$ \\
$\Sigma^{+}_c $&$\Sigma^{0}_b $& 1/2&  $1/{2}$ &  $I_K/2$ \\
$\Sigma^{0}_c $&$\Sigma^{-}_b $&$-1$ &$-m_\pi/m_K$   &$-I_\pi$ \\
$\Xi^{0'}_c $  &$\Xi^{-'}_b $  &$-1$&$-(1+m_\pi/ m_K)/2$ & $-(I_\pi+I_K)/2$ \\
$\Xi^{+'}_c $  &$\Xi^{0'}_b $  & 1/2&${m_\pi}/{(2 m_K)}$ &$I_\pi/2$  \\
$\Omega^{0}_c$ &$\Omega^{-}_b$ &$-1$&  $ -1$ &  $-I_K$\\
\hl
\end{tabular} 
\caption{Contributions to magnetic moments 
 of spin 1/2 c and b-baryons ($s_l=1$).  }
\label{tab:uno}
\end{center} 
\end{table}

The values of  the $\mu_i(B)$  contributions are reported in  Table~\ref{tab:uno}
 for baryons containing a  $Q$-quark ($Q=c,b$)
where
\bea\label{eq:intr}
I_i\equiv I(\Delta_{ST}, m_i) &=& 2 
\left(-2 +\log{\fr{m_i^{2}}{\mu^{2}}}
\right)+
2 {\sqrt{\Delta_{ST}^{2}- m_i^{2}}\over \Delta_{ST}}
\log{\left(\fr{\Delta_{ST}+ \sqrt{\Delta_{ST}^{2}- m_i^{2}} }{ \Delta_{ST}- 
\sqrt{\Delta_{ST}^{2}- m_i^{2}}
 }\right)} \ .
\eea

We want to stress that due to flavour symmetry, the constants $c_{s}$,
$g_{2}$ and $g_{3}$, and hence the values of  $\mu_{\chi}$, $\mu_{g_{2}}$ and $\mu_{g_{3}}$,
are the same for $c$ and $b$--baryons. The only difference is the contribution 
proportional to $\mu_{HQE}$ due to the different electric charge of the $c$, 
$Q_{c}=+2/3$, and $b$, $Q_{b}=-1/3$, quarks (see Eq.~\ref{eq:magS}).

In Eq.~\ref{eq:magS}  we have not considered
 contributions of order ${\cal O}(1/(m_{Q}\Lambda_{\chi}))$. 
 For the b--baryons, these 
corrections can be safely neglected. For the c--baryons, however, a simple 
estimate shows that their con\-tri\-bu\-tion cannot be larger than, say, 15\%.
Besides,
 the self--energy and loop diagrams with an insertion of the
operator in Eq. \ref{eq:long} yield contributions of order 
$O(1/\Lambda^{3}_{\chi})$ which again can be neglected because they are NNLO 
chiral corrections.

 The  results  proportional to $g_2^{2}$ are obtained performing a 
one-loop integral (fig.~\ref{fig:S} with an $S$--baryon running in the loop) 
that has to be renormalized.
The divergent part of the integral does not depend on  the pion or kaon mass
and is instead proportional to the  mass of the baryon running in the loop.
If one considers both pion and kaon loops the divergent part  respects
 the $SU(3)$ structure of the  chiral multiplet and  can be canceled
 with an operator of the  form
\beq
\frac{e}{\Lambda_{\chi}^2}
    {\rm tr }\left[\bar{S}_{\mu}\left(v\cdot D S_{\nu}\right) Q
 -\left( v\cdot D\bar{S}_{\mu}\right) S_{\nu} Q
\right] F^{\mu \nu}  \ . 
\label{eq:dseis}
\eeq
This  is the most general dimension--6 chiral and Lorentz 
invariant operator 
constructed out of $S_{\mu}^{i j}$ and $Q\, F_{\mu \nu}$, preserving parity 
and time-reversal invariance
which contributes to MM.
When the equation of motion 
($(v\cdot D)\, S_{\mu}\,=\, \Delta_{ST}\, S_{\mu}$) is applied,
 its contribution
 is of the same form as the term proportional to $c_{s}$ in
Eq.~\ref{eq:long}.
Thus, the local contribution from the operator in Eq.~\ref{eq:dseis}
 can be taken into account, together 
 with the  lowest order term in Eq.~\ref{eq:long},
through an effective coupling $c_S(\mu)$.
The scale $\mu$ dependence of the loop integrals  is exactly canceled
by the corresponding dependence of the coefficient
$c_S(\mu)$.

The contribution proportional to $g_3^{2}$ involves  a loop integral
 in which a baryon of the $T$-multiplet is running in the loop.
However,
 as we are in the limit of $m_T\rightarrow \infty$ no mass term
 for these $T$-baryons is present in the Lagrangian of   Eq.~\ref{eq:lagos}.
This means that the only massive particles running in the loop are 
the light mesons and
 the result of the integral is  convergent and proportional to  
 their mass.

 Using Table~\ref{tab:uno} one  can derive  the following
 linearly  independent relations for  the magnetic moments of spin 1/2
baryons containing a $c$-quark:
\bea
\mu(\S^{++}_c)+\mu(\S^{0}_c)&=&2\mu(\S^{+}_c)  \nn 
\\
\mu(\S^{++}_c)+\mu(\Omega^{0}_c)&=&2 \mu(\Xi^{+'}_c) \nn \\
\mu(\S^{++}_c)+2\mu(\Xi^{0'}_c)&=&\mu(\S^{0}_c)+2\mu(\Xi^{+'}_c) \nn \\
\mu(\S^{0}_c)+2\mu(\Xi^{+'}_c)&=&\fr{1}{6 m_c}\ . \label{eq:prel}
\eea
Including the spin 3/2 baryons  one can derive six more independent
relations,
\bea
\mu(\S^{++*}_c)+\mu(\S^{0*}_c)&=&2\mu(\S^{+*}_c)  \nn
\\
\mu(\S^{++*}_c)+\mu(\Omega^{0*}_c)&=&2 \mu(\Xi^{+'*}_c) \nn \\
\mu(\S^{++*}_c)+2\mu(\Xi^{0'*}_c)&=&\mu(\S^{0*}_c)+2\mu(\Xi^{+'*}_c)\nn \\
\mu(\S^{0*}_{_c })+2\mu(\Xi^{0'*}_{_c })&=&3\left(\mu(\S^{0}_c)+2\mu(\Xi^{+'}_c)\right) \nn \\
 \fr{2}{3}\mu(\S^{++*}_{c })&=& \mu(\S^{0}_c)+2\mu(\Xi^{+'}_c)-\mu(\S^{++}_c) \nn\\
6\mu(\S^{+}_c)-4\mu(\S^{++}_c)&=&  -4\mu(\S^{+*}_c)+\fr{8}{3}\mu(\S^{++*}_c) \ .
\label{eq:prel*}
\eea
The last three equations connect observables  corresponding to
spin 1/2  and  spin 3/2 baryons.

Moreover, it is easy
to deduce  10 analogous equations that  relate
baryons having a $b$-quark
\bea
\mu(\S^{+}_b)+\mu(\S^{-}_b)&=&2\mu(\S^{0}_b)  \nn \\
\mu(\S^{+}_b)+\mu(\Omega^{-}_b)&=&2 \mu(\Xi^{0'}_b)  \nn \\
\mu(\S^{+}_b)+2\mu(\Xi^{-'}_b)&=&\mu(\S^{-}_b)+2\mu(\Xi^{0'}_b) \nn  \\
\mu(\S^{-}_b)+2\mu(\Xi^{0'}_b)&=&-\fr{1}{12 m_b}  \nn \\
\mu(\S^{+*}_b)+\mu(\S^{-*}_b)&=&2\mu(\S^{0*}_b)  \nn \\
\mu(\S^{+*}_b)+\mu(\Omega^{-*}_b)&=&2 \mu(\Xi^{0'*}_b) \nn \\
\mu(\S^{+*}_b)+2\mu(\Xi^{-'*}_b)&=&\mu(\S^{-*}_b)+2\mu(\Xi^{0'*}_b) \nn \\
\mu(\S^{-*}_{_b })+2\mu(\Xi^{-'*}_{_b })&=&3\left(\mu(\S^{-}_b)+2\mu(\Xi^{0'}_b)\right) \nn
 \\
 \fr{2}{3}\mu(\S^{+*}_{b })&=& \mu(\S^{-}_b)+2\mu(\Xi^{0'}_b)-\mu(\S^{+}_b) \nn\\
6\mu(\S^{0}_b)-4\mu(\S^{+}_b)&=&
-4\mu(\S^{0*}_b)+\fr{8}{3}\mu(\S^{+*}_b) \ ,
\label{eq:prelb*}
\eea
 and two independent  equations that relate $b$-
 and $c$-baryons
\bea
\mu(\S^{0}_b)-\mu(\S^{+}_b)&=&\mu(\S^{+}_c)-\mu(\S^{++}_c) \nn \\
\mu(\S^{++}_c)-\fr{1}{3}\mu(\S^{++*}_{c })&=& \mu(\S^{+}_b)-\fr{1}{3}\mu(\S^{+*}_{b }) \ .
\eea

From Table~\ref{tab:uno}, we see that the order
 ${\cal O}(1/\Lambda_{\chi})$ and ${\cal O}(1/\Lambda_{\chi}^2)$
 contributions  cancel in the sum of all baryon MM within the sextet.
Therefore,  the average
over the baryon moments  measures the MM of the heavy quark,
\beq
\langle \mu(S_Q)\rangle =\fr{1}{3}\langle \mu(S_Q^*)\rangle=\fr{Q_Q}{12 m_Q}
\ .
\eeq

Notice also that  we can construct other combinations
 such that  
$c_{s}$, $g_{2}^{2}$ and $g_{3}^{2}$ contributions  cancel
\bea
\mu(\S^{++}_c)+\mu(\S^{0}_c)+\mu(\Omega^{0}_c) = 
\mu(\S^{+}_c)+\mu(\Xi^{0'}_c)+\mu(\Xi^{+'}_c) &=&
\frac{1}{6}\, \frac{\mu_{HQE}(B^{(*)})}{m_{c}} \nn
\\
\mu(\S^{+}_b)+\mu(\S^{-}_b)+\mu(\Omega^{-}_b) = 
\mu(\S^{0}_b)+\mu(\Xi^{-'}_b)+\mu(\Xi^{0'}_b)&=&
-\frac{1}{12}\, \frac{\mu_{HQE}(B^{(*)})}{m_{b}} \ .
\eea

If one has  got a numerical estimate of the couplings $g_2$ and $g_3$,
  it is possible to derive  a  scale independent relation  between any couple 
of baryons.
The combination 
\beq
\mu(B_1) - \fr{\mu_\chi(B_1)}{\mu_\chi(B_2)}\mu(B_2)
\eeq
is independent of 
 the unknown coupling $c_S(\mu)$  an can then be predicted.
For instance
\bea
\mu(\S^{+}_b)+2 \mu(\S^{-}_b)&=& \fr{1}{24}
\fr{g_3^{2}}{4 \pi f_\pi^{2}} (m_K-m_\pi)-
               \fr{\Delta_{ST}}{24}\fr{g_2^{2}}{(4 \pi f_\pi)^{2}} 
(I_K-I_\pi) - \fr{1}{12 m_b}\ .
\label{eq:dif}
\eea
The couplings $g_2$ and  $g_3$ 
 have been calculated theoretically. In Table~\ref{tab:gteor}
 we report   the results of these computations.

There exists  an  experimental measurement of $g_3$  from CLEO
 coming from the  decay $\S_c^*\rightarrow \Lambda_c \pi $~\cite{cleo,gnec},
$g_3=\sqrt{3} \,(0.57\pm 0.10)$.
The direct measurement of $g_2$ is not possible at present.
However, the quark model relates its value to $g_3$~\cite{gnec}, yielding
$g_2=1.40 \pm 0.25$.

\begin{table}
\begin{center}
\begin{tabular}{|c|c|c|} \hl
Model &
$g_2$ & $g_3$ \\ \hl
Large $N_c$ ~\cite{gural} & 1.88 & 1.53\\
Quark model ~\cite{gnuc} & $1.5$ & $1.06 $ \\
Short distance QCD sum rule ~\cite{grozi} & $0.83 \pm 0.23$ & $0.67 \pm 0.18$\\
Light-cone  QCD sum rules ~\cite{zhu} & $1.56\pm 0.3\pm 0.3$ &
$0.94 \pm 0.06\pm 0.2$ \\ \hl
\end{tabular}
\caption{Theoretical estimates of $g_2$ and $g_3$.}
\label{tab:gteor}
\end{center}
\end{table} 

In order to get a numerical estimate of the left-hand side of Eq.~\ref{eq:dif}
we set $g_2=1.5 \pm 0.3$ and $g_3=0.99 \pm 0.17$ 
and the rest of the constants
 as in Table~\ref{tab:cost}.
We find 
for our best estimate of  Eq.\ref{eq:dif}
\beq
\mu(\S^{+}_b)+2\mu(\S^{-}_b)=0.23 \pm 0.09 \ {\rm GeV}^{-1} \ . 
\eeq

\section{Results for $T$-baryons ($s_l=0$)}
\label{sec:T}

As the light quarks of the $T$-baryons are  in a $s_l=0$ configuration,
  the contributions to the magnetic moments of these hadrons  are $1/m_Q$  suppressed~\cite{cho}. 
The leading term  is of the form $\mu_{HQE}/m_Q$ and  the  first chiral corrections are
 of order  ${\cal O}(1/(m_{Q}\Lambda_{\chi}))$  and  come from~\cite{sava}
\beq
{\cal{L}}^{\prime}_{(long)} = \fr{c_T}{4 m_Q} \fr{e}{\Lambda_\chi}
{\bar T}^i \sigma_{\mu\nu} Q_{ij} T^j F^{\mu\nu} \ .
\eeq
The contributions of  order ${\cal O}(1/(m_{Q}\Lambda_{\chi}^2))$  have different origin:
\begin{enumerate}
\item
there is 
a divergent contribution~\cite{sava} coming  from Eq.~\ref{eq:lagos}
 through the chiral loops shown in fig. 2, which is proportional to the 
 explicit mass splitting,
\beq
 \Delta M_Q= 3\fr{\la_{2S}}{m_Q} \ ,
\eeq 
  for the  spin 1/2 and spin 3/2  parts of $S$-baryons~\cite{jenk};
 \item besides, one can consider
 a spin--symmetry breaking operator of $O(1/m_{Q})$; 
\beq
{\cal{L}}^{\prime} =
\fr{g'}{m_Q}
 \left [\epsilon_{ijk} \bar{T}^i\sigma^{\mu\nu} (\xi_{\mu})_l^j S_{\nu}^{kl}+
 \epsilon^{ijk} \bar{S}_{kl}^{\mu} \sigma_{\mu\nu} (\xi^{\nu})_j^l T_i\right ] \ ,
\label{eq:nova}
\eeq
which gives rise to divergent loop diagrams, as the one in fig. 2, where one
of the vertices is proportional to $g'$.
\item further, there are finite contributions of the same order coming from the 
$SU(3)$-breaking operators
 \beq
e \fr{\omega_1}{4 m_Q  \Lambda_\chi^2}
 {\bar T}^i \sigma_{\mu\nu} Q_{il} \chi^l_j T^j F^{\mu\nu} +
e \fr{\omega_2}{4 m_Q  \Lambda_\chi^2} Q_Q
 {\bar T}^i \sigma_{\mu\nu} \chi_{ij} T^j F^{\mu\nu} \ ,
\label{eq:nova2}
\eeq
where, in the limit of exact isospin symmetry
\beq
 \chi=\left(
\begin{array}{ccc}
m_\pi^2 && \\
&m_\pi^2 & \\ 
&&2 m_K^2- m_\pi^2  
\end{array}
 \right) \ .
\eeq
\end{enumerate}
 As in the case of the $S$-baryons, when all Goldstone boson loops are included,
 the scale $\mu$ dependence of the result of fig. 2 
 is canceled by  the   corresponding  dependence  of an effective  $c_T(\mu)$. 
Neither the  interaction term  of Eq.~\ref{eq:nova}
nor the finite terms of Eq.~\ref{eq:nova2} 
were taken into account in ref.~\cite{sava}.

\begin{figure}[ht]
\begin{center}
\epsfig{file=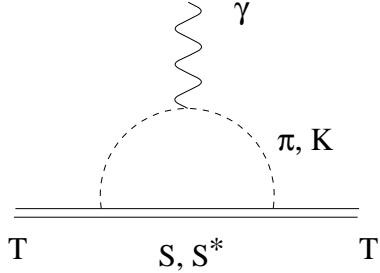,width=0.3\linewidth,angle=0}
\caption{Meson loops contributing to $T$-baryons MM.}
\label{fig:T}
\end{center}
\end{figure}

 Similarly to what we have done  in the previous paragraph we write
 the magnetic moment of $T$-baryons as
\bea
\mu (B)&=& \fr{1}{24 m_Q}\left(-6 Q_{Q}\mu_{HQE}(B)-
 \fr{ c_T}{ \Lambda_\chi }\mu_{T}(B)+
 g_3^{2} \fr{  3 \la_{2S}}{(4 \pi f_\pi)^2}\mu_{g_3}(B) \right.\nn \\
&& \left.
+6 g_3  g' \fr{\Delta_{ST}}{(4 \pi f_\pi)^2}\mu_{g'}(B) + 
2 \fr{\omega_1 m_K^2}{\Lambda_\chi^2}
\mu_{\chi_1} (B) -6 Q_{Q}
\fr{\omega_2 m_K^2}{\Lambda_\chi^2 }
\, \mu_{\chi_2} (B)
\right)\ .
\eea
 The values  of  the  $\mu_i$ are written in Table~\ref{tab:t} where
\beq
J_i=\fr{\partial}{ \partial \Delta_{ST}} 
\left( \Delta_{ST}I(\Delta_{ST}, m_i) \right) \ .
\eeq
Corrections 
to our results for
$T$--baryons  are of order ${\cal O}(1/m_{Q}^{2})$ and hence negligible.

By eliminating the unknown coupling constants,
one  can deduce two independent relations among the magnetic moments of both
T-multiplets
\bea
m_b\, \mu (\Xi_b^-)-m_c\, \mu (\Xi_c^0)& =&
m_b\, \mu (\Xi_b^0)-m_c\, \mu (\Xi_c^+) \\
m_b\, \mu (\Lambda_b^0)-m_c\, \mu (\Lambda_c^+)- \fr{1}{4} &=& 
\left( 2\fr{m_K^2}{m_\pi^2}-1\right) 
\left[ m_b\, \mu (\Xi_b^-)-m_c\, \mu (\Xi_c^0)-\fr{1}{4}\right]
\ .
\eea

In the absence of the $SU(3)$-breaking operators in Eq.~\ref{eq:nova2}, 
the average baryon MM over the $T$ multiplet would be
equal to the heavy quark MM~\cite{sava}.
The result is however corrected by contributions proportional 
to the unknown couplings $\omega_1$ and
 $\omega_2$\footnote{Notice that our definition of MM differs from the one
in ref.~\cite{sava} by a factor $-1/4$.}:
\beq
\langle \mu(T_Q)\rangle=
-\fr{1}{4 m_c} \left [ Q_Q
+ \fr{2 \omega_1 m_K^2}{9 \La_\chi^2}\left ( 1- \fr{m_\pi^2}{m_K^2}\right )
- Q_Q \fr{\omega_2 m_K^2}{3 \La_\chi^2}\left ( 2+ \fr{m_\pi^2}{m_K^2}\right )
\right ] \ .
\eeq

\begin{table}
\begin{center} 
\begin{tabular}{|c|c|c|c|c|c|c|}  \hl
 c quark & b quark & 
 $ \mu_T  $   &
$\mu_{g_3}   $ &
$\mu_{g'}   $ &
$\mu_{\chi_1}$ &
$\mu_{\chi_2}$ 
\\
\hl
$\Xi^{0}_c $  &$\Xi^{-}_b $  &$4$ &$J_\pi+J_K $ & $I_\pi+I_K$ & $-2m^2_\pi/m_K^2$ & $m^2_\pi/m_K^2$ \\
$\Xi^{+}_c $  &$\Xi^{0}_b $  &$-2$&$-J_\pi$ & $-I_\pi$ & $m^2_\pi/m_K^2$ & $m^2_\pi/m_K^2$ \\
$\Lambda_c^{+}$   &$\Lambda_b^{0}$   &$-2$&$-J_K$ &  $-I_K$ & $2-m^2_\pi/m_K^2$ & 
 $2-m^2_\pi/m_K^2$ \\
\hl
\end{tabular} 
\caption{Contributions to magnetic moments 
 of spin 1/2 $T$-baryons ($s_l=0$).  }
\label{tab:t}
\end{center} 
\end{table}

\section{Conclusions}
\label{sec:fin}
The magnetic moments of triplet and sextet heavy baryons have been computed
in the HHCPT.
The calculation    of the $S^{(*)}$-baryons MM  at the order ${\cal O}(1/\Lambda_\chi^2)$
 involves only one new arbitrary constant,
$c_S$.
 Thus it is possible  to derive  relations among the MM of the  hadrons in the same sextet
 where all masses and  effective couplings  are eliminated.
Due to heavy quark symmetry  the MM of the $S$ and  $S^*$ sextets  are also related.
Moreover, as $c$ and $b$  baryons are described by the same arbitrary constants,
we can connect  the MM of the two kinds of hadrons.
The average over  one sextet equals the corresponding heavy quark MM.

In the case of  $T$-baryons the first corrections appear at 
  order ${\cal O}(1/(m_Q \Lambda_\chi^2))$ and  four   arbitrary constants 
are required.
Then  we are left with only two independent relations  which combine
 $c$ and $b$ triplets  and contain $m_c$ and $m_b$.
The average over  one triplet equals the heavy quark MM only in  the absence
of $SU(3)$-breaking operators.

The  measure of the magnetic moments  of heavy baryons represents an
experimental challenge. Nevertheless  several groups are contemplating
the possibility of performing it  in the near future 
(BTeV, SELEX)~\cite{exp}.

\section*{Acknowledgements}
I.S. wants to thank A. Della Riccia Foundation (Florence, Italy) for support. 
M.C.B. is indebted to the Spanish Ministry of Education and Culture 
for her fellowship.
This work has been supported in part by the European Union TMR Network  
``EURODAPHNE'' (Contract No. ERBFMX-CT98-0169), by DGESIC, Spain
(Grant No. PB97-1261) and by CICYT (Grant No.  AEN-96/1718).

\end{document}